\documentclass[aps,prl,showpacs,twocolumn,superscriptaddress]{revtex4-1}
\usepackage{bm,color}
\usepackage{graphicx}
\usepackage{amsmath,amssymb}
\usepackage{braket}
\begin{document}
\title{Low-Energy Excitations of Skyrmion Crystals in a Centrosymmetric Kondo-Lattice Magnet: Decoupled Spin-Charge Excitations and Nonreciprocity}
\author{Rintaro Eto}
\affiliation
{Department of Applied Physics, Waseda University, Okubo, Shinjuku-ku, Tokyo 169-8555, Japan}
\author{Rico Pohle}
\affiliation
{Department of Applied Physics, Waseda University, Okubo, Shinjuku-ku, Tokyo 169-8555, Japan}
\affiliation
{Department of Applied Physics, University of Tokyo, Hongo, Bunkyo-ku, Tokyo 113-8656, Japan}
\author{Masahito Mochizuki}
\affiliation
{Department of Applied Physics, Waseda University, Okubo, Shinjuku-ku, Tokyo 169-8555, Japan}
\begin{abstract}
We theoretically study spin and charge excitations of skyrmion crystals stabilized by conduction-electron-mediated magnetic interactions via spin-charge coupling in a centrosymmetric Kondo-lattice model by large-scale spin-dynamics simulations combined with the kernel polynomial method. We reveal clear segregation of spin and charge excitation channels and nonreciprocal nature of the spin excitations governed by the Fermi-surface geometry, which are unique to the skyrmion crystals in centrosymmetric itinerant hosts and can be a source of novel physical phenomena.
\end{abstract}

\maketitle
Magnetic skyrmions~\cite{Bogdanov1989,Bogdanov1994,Rossler2006,Nagaosa2013,Mochizuki2015,Seki2016,Everschor2019,Tokura2021}, swirling magnetic textures characterized by a quantized topological invariant, have attracted great research interest since an experimental discovery of their crystallization into a skyrmion crystal (SkX) in chiral magnets~\cite{Muhlbauer2009,Yu2010}. The noncoplanar skyrmion magnetizations generate emergent magnetic fields acting on conduction electrons through giving them a Berry phase via exchange interactions, which give rise to novel transport phenomena~\cite{YeJ1999,Ohgushi2000,Bruno2004,Onoda2004,YiSD2009,ZangJ2011}, e.g., topological Hall effects~\cite{MLee2009,Neubauer2009,Schulz2012, Ritz2013,HuangSX2012} and topological Nernst effects~\cite{Shiomi2013,Mizuta2016,Mizuta2018}. The SkXs in the chiral magnets are stabilized by the Dzyaloshinskii-Moriya interaction (DMI) of relativistic origin~\cite{Dzyaloshinsky1958,Moriya1960a,Moriya1960b}, which becomes active in chiral crystal structures with broken spatial inversion symmetry.

Recently, emergence of SkX has been predicted also in centrosymmetric magnets, in which the DMI can no longer be a stabilization mechanism because of its absence. Instead, several new mechanisms have been proposed theoretically~\cite{Batista2016}. One important mechanism is frustrated exchange interactions due to lattice geometry~\cite{Okubo2012,Kamiya2014,Leonov2015,LinSZ2016,Hayami2016a,Hayami2016b,Kharkov2017,LinSZ2018,Lohani2019,GaoS2020,WangZ2021,Aoyama2022}. Another important mechanism is indirect spin interactions mediated by conduction electrons in itinerant Kondo-lattice magnets~\cite{Ruderman1954,Kasuya1956,Yosida1957}. In the Kondo-lattice magnets, the SkXs emerge as superpositions of multiple magnetic helices whose propagation vectors $\bm Q_i$ are determined by nestings of Fermi surface~\cite{Hayami2021R,Ozawa2016,Ozawa2017,Hayami2017,Gobel2017,WangZ2020,Hayami2020a,Hayami2021a,Hayami2021b,WangZ2022}. Subsequent experiments have successively discovered several centrosymmetric itinerant magnets hosting SkXs possibly stabilized by the latter mechanism, e.g., Gd$_2$PdSi$_3$~\cite{Kurumaji2019,Hirschberger2020a,Hirschberger2020b,Nomoto2020}, Gd$_3$Ru$_4$Al$_{12}$~\cite{Hirschberger2019,Hirschberger2021}, and GdRu$_2$Si$_2$~\cite{Khanh2020,Yasui2020,Khanh2022}.

Skyrmions in these newly discovered magnets possess a distinctive feature, that is, several degrees of freedom such as chirality, helicity, and vorticity survive in the absence of DMI as opposed to the skyrmions in chiral magnets. However, these unfrozen degrees of freedom scarcely appear at equilibrium and even in transport properties~\cite{Kurumaji2019,Hirschberger2020a}. This is because although these degrees of freedom remain in high-temperature disordered phases, they are frozen in ordered phases at lower temperatures through spontaneous symmetry breaking upon phase transitions. Instead of equilibrium and transport properties, the unique properties of SkXs in these magnets are expected to show up in dynamical phenomena~\cite{Leonov2015,Leonov2017,ZhangX2017,YaoX2020,Eto2021}. For example, possible microwave-induced switching of magnetic topology has been theoretically predicted~\cite{Eto2021}. Furthermore, the physics of skyrmion crystals in centrosymmetric systems shares many similarities with other research fields, e.g., elementary particles~\cite{Skyrme1961}, liquid crystals~\cite{Wright1989,Matteis2020,Matteis2022}, Bose-Einstein condensates~\cite{Khawaja2001}, quantum Hall magnets~\cite{Sondhi1993,Brey1995}, vortex lattices in type II superconductors~\cite{Abrikosov2004}, and even quantum computing technologies~\cite{Psaroudaki2021,Jia2022}. Under these circumstances, clarification of spin and charge excitations of SkXs in the centrosymmetric magnets are of crucial importance because it will inevitably open a new horizon of research on topologies in matters.

In this Letter, we theoretically study the spin and charge excitations of SkXs in the centrosymmetric Kondo-lattice model. We reveal that the SkX at zero field has three linearly dispersive Goldstone modes associated with the breaking of SO(3) symmetry in the spin space and one quadratically dispersive pseudo-Goldstone mode associated with the breaking of translational symmetry. Surprisingly we find clear segregation of spin and charge channels in these excitations. We also uncover that when a magnetic field is applied, some of the Goldstone modes become gapped, and the spin excitations attain nonreciprocal nature caused by induced asymmetries of the Fermi surface. These rich properties of spin and charge excitations are key attributes of skyrmions in the centrosymmetric spin-charge coupled magnets governed by the Fermi-surface geometry.

We consider the Kondo-lattice model on a triangular lattice as a typical model of the centrosymmetric itinerant magnets with spin-charge coupling,
\begin{multline}
\mathcal{H}_{\rm KLM}=
-\sum_{ij\sigma}t_{ij} \hat{c}^\dagger_{i\sigma}\hat{c}_{j\sigma}
-\mu \sum_{i \sigma} \hat{c}^\dagger_{i\sigma}\hat{c}_{i\sigma} \\
-J_{\rm K}\sum_i \hat{\bm s}_i \cdot {\bm S}_i
-H_z\sum_i S_{i}^z,
\label{eq:HamKLM}
\end{multline}
where $\hat{c}^\dagger_{i\sigma}$ ($\hat{c}_{i\sigma}$) denotes a creation (annihilation) operator of a conduction electron with spin $\sigma(=\uparrow,\downarrow)$ on site $i$. The first and second terms describe kinetic energies and chemical potential of conduction electrons with the nearest-neighbor hopping $t_1$(=1), the third-nearest-neighbor hopping $t_3$(=$-0.85$), and the chemical potential $\mu$(=$-3.5$). The third term describes the Kondo exchange coupling between the localized classical spins $\bm S_i$ ($|\bm S_i|=1$) and the conduction-electron spins $\hat{\bm s}_i=\frac{1}{2}\sum_{\sigma\sigma^\prime}\hat{c}^\dagger_{i\sigma}{\bm \sigma}_{\sigma\sigma^\prime}\hat{c}_{i\sigma^\prime}$ where $\bm \sigma$ is the vector of Pauli matrices. The fourth term is the Zeeman coupling term associated with an external magnetic field $\bm H=(0,0,H_z)$ perpendicular to the lattice plane. Ozawa and coworkers studied this model when $J_{\rm K}=1$ and found emergence of three phases argued below as a function of $H_z$~\cite{Ozawa2017}. Note that the exchange gap is not opened for the present weak-coupling case with $J_{\rm K}\lesssim2t_1$. The chemical potential is tuned to reproduce the experimentally observed skyrmion size of $\sim$3 nm in Gd$_2$PdSi$_3$ with magnetic ordering vectors of $|\bm Q_\nu|=\pi/3$~\cite{Kurumaji2019}, but we have confirmed that the results do not depend on its choice so much.

\begin{figure}[tb]
\centering
\includegraphics[scale=0.5]{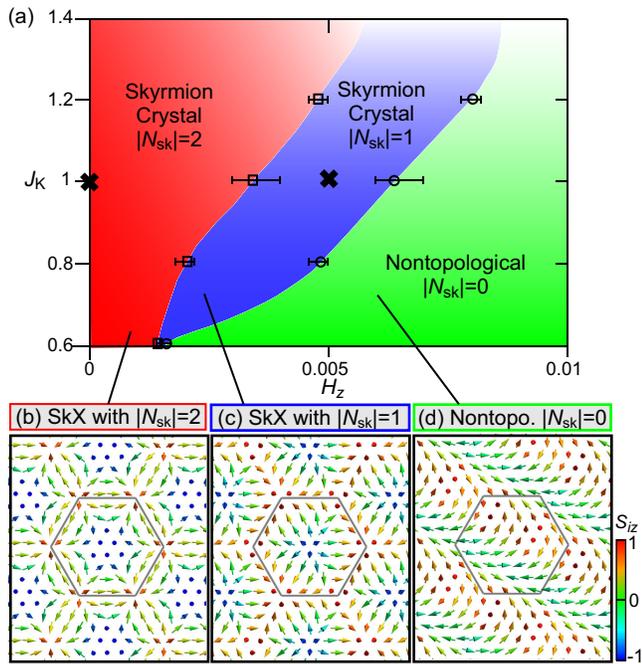}
\caption{(a) Ground-state phase diagram in plane of $H_z$ and $J_{\rm K}$ for the Kondo-lattice model in Eq.~(\ref{eq:HamKLM}). (b)-(d) Spatial spin configurations of (b) SkX with $|N_{\rm sk}|=2$, (c) SkX with $|N_{\rm sk}|=1$, and (d) nontopological state with $N_{\rm sk}=0$. Magnetic unit cells of these phases (gray hexagons) contain 48 sites. The excitation spectra in Fig.~\ref{Fig02} and Fig.~\ref{Fig03} are calculated at the points indicated by cross symbols in (a).}
\label{Fig01}
\end{figure}
We first calculate the ground-state phase diagram [Fig.~\ref{Fig01}(a)] and the spin configurations [Figs.~\ref{Fig01}(b)-(d)] of this Hamiltonian by using the kernel polynomial method~\cite{Wang1994,Motome1999,Weisse2006,Tang2012} combined with the over-damped spin dynamics simulation~\cite{Barros2013,Wang2016,Ozawa2017b,Wang2018} for a triangular-lattice system of $N=96^2$ sites with periodic boundary conditions. Details of the numerical calculations are described in Supplemental Material~\cite{SM}.
The obtained phase diagram contains three phases, i.e., a SkX phase with $|N_{\rm sk}|=2$, another SkX phase with $|N_{\rm sk}|=1$, and a nontopological phase with $N_{\rm sk}=0$. When the value of $J_{\rm K}$ is fixed, these phases emerge in this order as $H_z$ increases. Here $N_{\rm sk}$ is a topological invariant called skyrmion number, which characterizes a topology of each magnetic state described by a superposition of three spin-density waves.

To calculate the spin and charge excitation spectra in each phase, we perform the quantum Landau-Lifshitz dynamics simulation, which is known to be powerful to study the dynamical phenomena of spin-charge coupled systems~\cite{Chern2018}. Here we assume that the itinerant electrons smoothly follow the dynamics of localized spins being immediately relaxed to the ground state for a given localized-spin configuration at each moment. Within this adiabatic approximation, we formulate an effective magnetic field $\bm H_i^{\rm eff}$ that acts on the localized spins as $\bm H_i^{\rm eff}=-\partial \Omega/\partial \bm S_i$, where $\Omega$ is the thermodynamical potential. We calculate $\Omega$ and ${\bm H}_i^{\rm eff}$ using the kernel polynomial method (see Supplemental Material for details). The effective field ${\bm H}_i^{\rm eff}$ induces dynamics of localized spins which obey a generalized form of the Landau-Lifshitz equation,
\begin{equation}
\frac{d\bm S_i}{dt} = \frac{i}{\hbar}[\bm S_i, \mathcal{H}_{\rm KLM}]
\approx \bm H_i^{\rm eff} \times \bm S_i.
\label{EoMKLM}
\end{equation}
We numerically solve this equation using the fourth-order Runge-Kutta method to trace time-space evolutions of the localized spins $\bm S_i(t)$ after locally applying a short magnetic-field pulse~\cite{Mochizuki2015,Mochizuki2012,Mochizuki2010}. The wavefunction of conduction electrons $\ket{\Psi(t)}$ at each moment $t$ is composed of eigenstates $\ket{\psi_\nu(t)}$ of the Hamiltonian $\mathcal{H}_{\rm KLM}$.

\begin{figure*}[tbh]
\centering
\includegraphics[scale=1.0]{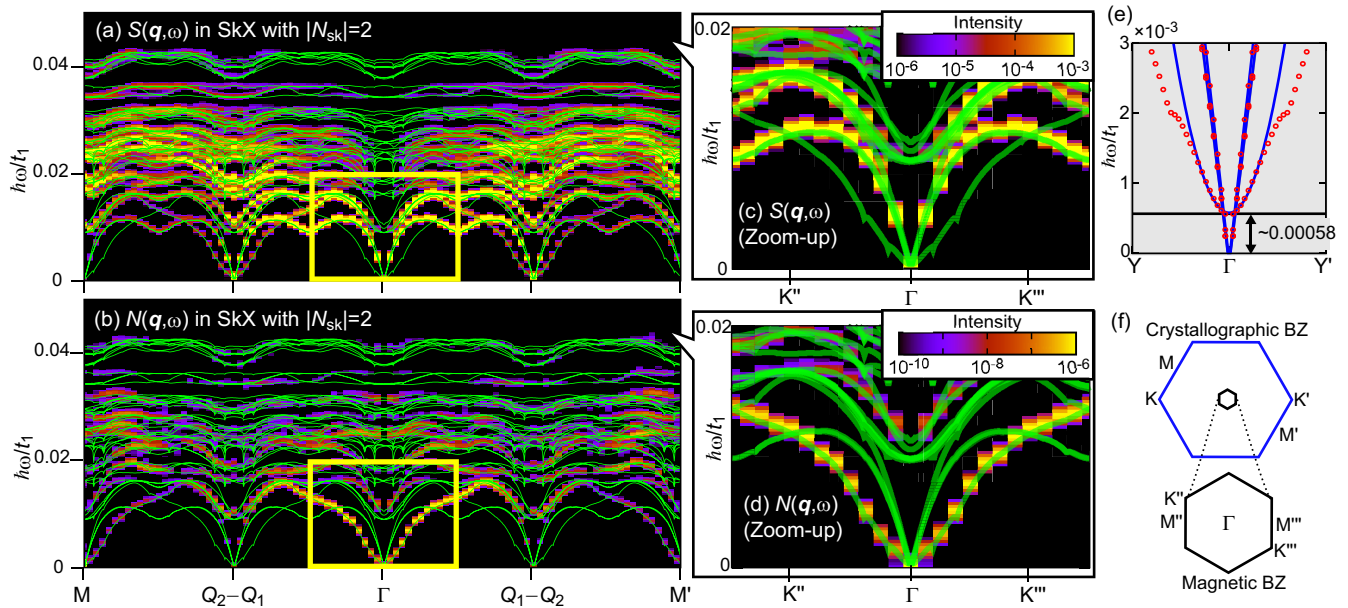}
\caption{Results for the SkX with $|N_{\rm sk}|=2$ when $J_{\rm K}=1$ and $H_z=0$. (a),~(b) Dynamical spin and charge structure factors, (a) $S(\bm q,\omega)$ and (b) $N(\bm q,\omega)$, in the momentum-frequency plane. (c),~(d) Zoom-up on the area around the $\Gamma$ point indicated by rectangles in (a) and (b), respectively. The spectra calculated by the spin dynamics simulations are presented by colors, while the dispersion relations obtained by the linear spin-wave theory are presented by thick (green) lines. (e) Magnified dispersion relations of the spin excitation near the $\Gamma$ point with data calculated by the linear spin-wave theory (red points) and fitting curves. The Y and Y$^\prime$ points correspond to momenta $(-\sqrt{3}\pi/100, \pi/100)$ and $(\sqrt{3}\pi/100, -\pi/100)$, respectively. (f) Crystallographic (blue) and magnetic (black) Brillouin zones.}
\label{Fig02}
\end{figure*}
Dynamical spin and charge structure factors, $S(\bm q,\omega)$ and $N(\bm q,\omega)$, are obtained from the simulated time profiles of $\bm S_i(t)$ and $\ket{\Psi(t)}$ as,
\begin{eqnarray}
S(\bm q,\omega)=\frac{1}{\sqrt{N_t} N}\sum_{i,j}^N e^{i\bm q \cdot (\bm r_i-\bm r_j)}
\sum_{n=1}^{N_t} e^{i \omega t_n} \braket{\bm S_i(t) \cdot \bm S_j(0)},
\nonumber
\label{eq:Sqw}
\\
N(\bm q,\omega)=\frac{1}{\sqrt{N_t} N}\sum_{i,j}^N e^{i\bm q \cdot (\bm r_i-\bm r_j)}
\sum_{n=1}^{N_t} e^{i \omega t_n} \braket{n_i(t) \cdot n_j(0)},
\nonumber
\label{eq:Nqw}
\end{eqnarray}
where $n_i(t)=\sum_\sigma\bra{\Psi(t)}\hat{c}^\dagger_{i\sigma}\hat{c}_{i\sigma}\ket{\Psi(t)}$ is the expectation value of the electron number on site $i$ at time $t$. We also calculate dispersion relations of the excitation spectra by extending a framework of the linear spin-wave theory for the Kondo-lattice model in Refs.~\cite{Akagi2013,Akagi2014}.

Figures~\ref{Fig02}(a)-(e) show the calculated dynamical spin and charge structure factors, $S(\bm q,\omega)$ and $N(\bm q,\omega)$, for the SkX phase with $|N_{\rm sk}|=2$ when $J_{\rm K}=1$ and $H_z=0$. Since the magnetic unit cell contains 48 sites, each spectrum is composed of 48 bands within the total bandwidth of $\hbar\omega\approx0.04t_1$. The crystallographic and magnetic Brillouin zones are presented in Fig.~\ref{Fig02}(f). We find that the spectra calculated by the spin dynamics simulations (colors) and the dispersion relations calculated by the linear spin-wave theory (thick green lines) perfectly coincide with each other over the entire momentum-frequency region. 

Figures.~\ref{Fig02}(c) and (d) magnify areas around the $\Gamma$ point in Figs.~\ref{Fig02}(a) and (b), respectively.
In Fig.~\ref{Fig02}(c), the spin-excitation spectrum of $S(\bm q,\omega)$ has large intensities on the three linearly dispersive gapless modes (Goldstone modes), while no intensity is observed on the quadratically dispersive mode. These three Goldstone modes are associated with three generators of the SO(3) group in the spin space and originates from spontaneous breaking of the SO(3) symmetry. 

On the contrary, the charge-excitation spectrum of $N({\bm q},\omega)$ in Fig.~\ref{Fig02}(d) has a large spectral intensity on the quadratic mode, whereas no intensity is observed on the three linearly dispersive Goldstone modes. The quadratic mode is associated with a translational symmetry. In an exact sense, the translational symmetry in this system is discrete, which gives rise to a small spin-wave gap. However, it can be regarded as a nearly continuous one because a spatial period of the SkX is much longer than the lattice constants, for which the gap becomes extremely small. Indeed, a magnitude of the gap is evaluated to be $\Delta\sim0.00058t_1$ through fitting the data with a function $\hbar\omega/t_1=vq^2+\Delta$ [Fig.~\ref{Fig02}(e)]. The gap is evaluated to be less than 1.5\% of the full spectral bandwidth. This nearly gapless quadratic mode originates from spontaneous breaking of the pseudo continuous translational symmetry and, thus, can be regarded as a pseudo-Goldstone mode~\cite{Weinberg1972,Burgess2000}.

\begin{figure*}[tbh]
\centering
\includegraphics[scale=1.0]{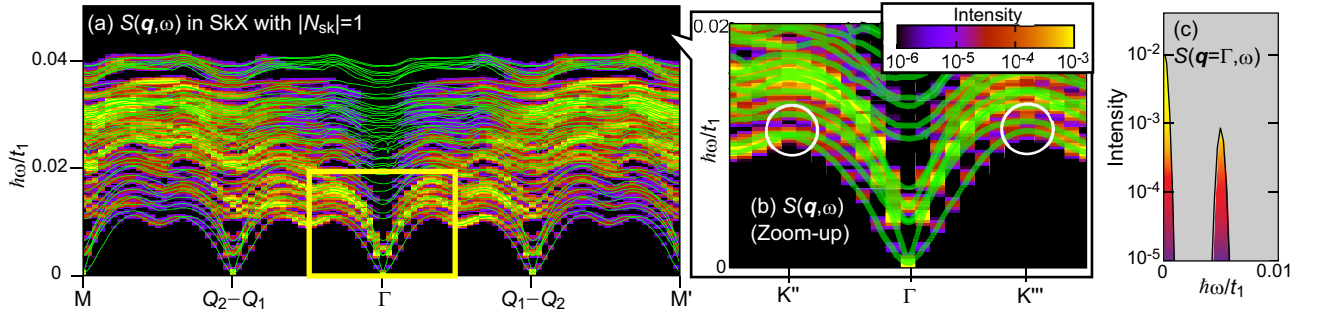}
\caption{Results for the SkX with $|N_{\rm sk}|=1$ when $J_{\rm K}=1$ and $H_z=0.005$. (a) Dynamical spin structure factor $S(\bm q,\omega)$. (b) Zoom-up on the area indicated by a rectangle in (a). The spectra calculated by the spin dynamics simulations are presented by colors, whereas the dispersion relations obtained by the linear spin-wave theory are presented by thick (green) lines. Nonreciprocal nature of the spin-wave propagation appears in the dispersion relations at the momenta indicated by circles. (c) Momentum line-cut of $S(\bm q=\Gamma,\omega)$. The mode at $\hbar\omega=0.005t_1$ is the counterclockwise rotational mode.}
\label{Fig03}
\end{figure*}
It is surprising that the spin and charge excitations appear in different modes. This clear segregation is attributable to a fact that the global SO(3) spin rotation and the global translation of skyrmion-crystal configuration are totally independent and thus are decoupled, which might enable us to selectively observe and activate the magnetic and charge dynamics of SkX in centrosymmetric itinerant magnets by, e.g., inelastic neutron-scattering and resonant inelastic X-ray scattering experiments. However, this does not mean that the spins and charges are totally decoupled in excitations. The effect of spin-charge coupling shows up in the excitations of scalar spin chiralities. We have found that the dynamical structure factor of the scalar spin chirality $C_\chi(\bm q,\omega)$ (shown in Supplemental Material) has dominant spectral weight on the quadratic mode similar to the charge structure factor $N(\bm q,\omega)$. Moreover, they coincide nearly perfectly in intensity upon multiplying a scaling factor. The scalar spin chirality $\chi_i$ is defined by a solid angle spanned by three neighboring spins as $\chi_i=\bm S_i \cdot (\bm S_{i+\hat{\bm a}} \times \bm S_{i+\hat{\bm b}})/4\pi$ ($\hat{\bm a}$ and $\hat{\bm b}$ are the primitive translation vectors of triangular lattice) and is related with the skyrmion number as $N_{\rm sk}=N_{\rm muc} \sum_i \chi_i/N$ with $N_{\rm muc}(=48)$ being the number of sites in a magnetic unit cell. The dynamics of noncollinear spin texture with nonzero scalar spin chirality can induce the charge dynamics through generating the emergent electromagnetic field that acts on the conduction electrons via the Berry phase mechanism~\cite{Bruno2004,Chern2018}. The perfect coincidence of charge and chirality excitations can be understood by this mutual coupling.

We next discuss results for the SkX phase with $|N_{\rm sk}|=1$ when $J_{\rm K}=1$ and $H_z=0.005$. The spin-excitation spectrum $S(\bm q,\omega)$ in Fig.~\ref{Fig03}(a) is again composed of 48 bands within the total bandwidth of $\hbar\omega \approx 0.04t_1$, but the bands are less degenerate under application of external magnetic field. According to a closer look into the long-wavelength region [Fig.~\ref{Fig03}(b)], there exists only one linearly dispersive Goldstone mode associated with the  spontaneously broken U(1) symmetry of the localized spins in the presence of magnetic field. Specifically, the applied magnetic field reduces the SO(3) symmetry to the U(1) symmetry, where two of the three linearly dispersive Goldstone modes at zero field become gapped. The spectrum again shows a quadratic mode with an extremely tiny gap associated with the spontaneous breaking of the nearly continuous translational symmetry. We also find that the applied magnetic field causes mixing of spin and charge excitations so that their segregation observed in the SkX with $|N_{\rm sk}|=2$ becomes obscure in the SkX with $|N_{\rm sk}|=1$.

A momentum line-cut of $S(\bm q=\Gamma,\omega)$ in Fig.~\ref{Fig03}(c) shows that there exists a mode at finite frequency of $\hbar\omega=0.005t_1$. This mode is identified as a counterclockwise rotation mode, in which skyrmions in the SkX uniformly rotate in a counterclockwise fashion~\cite{Mochizuki2012,Leonov2015,Eto2021}. This mode originates from precessions of spins constituting the SkX around the magnetic field $H_z$, and its resonance frequency $\omega_{\rm R}$ is determined by the strength of $H_z$ as $\hbar\omega_{\rm R}/t_1=H_z(=0.005t_1)$. Note that SkXs in chiral magnets with DMI exhibit a pair of low-lying modes, i.e., a lower-frequency counterclockwise rotation mode and a higher-frequency clockwise rotation mode~\cite{Mochizuki2012}. On the contrary, the SkX in the present centrosymmetric system exhibits only the counterclockwise mode, and the clockwise mode is absent~\cite{Leonov2015,Eto2021}.

\begin{figure}[tbh]
\centering
\includegraphics[scale=0.5]{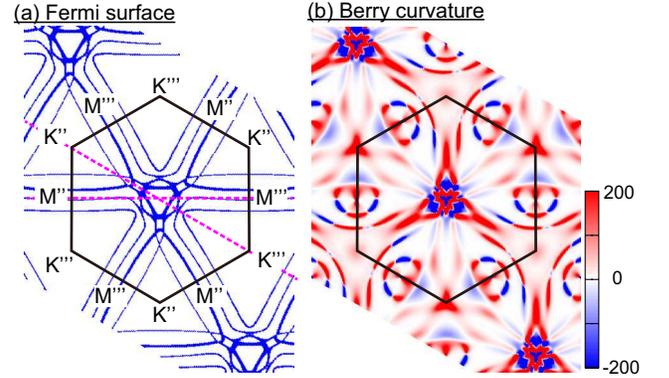}
\caption{Nonreciprocal nature of the SkX with $|N_{\rm sk}|=1$ shows up in the momentum-space electronic structure, i.e., (a) Fermi surface and (b) Berry curvature $B^z_m(\bm q)$ for the seventh band ($m$=7) of conduction electrons. The hexagons indicate the magnetic Brillouin zone.}
\label{Fig04}
\end{figure}
A remarkable property of spin excitation in the SkX with $|N_{\rm sk}|=1$ is the nonreciprocity, i.e., $S(\bm q,\omega) \ne S(-\bm q,\omega)$, seen in comparison between the spin-wave dispersions indicated by two circles in Fig.~\ref{Fig03}(b). This effect is caused by the symmetry of SkX spin structure that governs the symmetry of electronic structure. This aspect can be seen in the Fermi surface and the Berry curvature $B^z_m(\bm q)$. The Berry curvature is calculated by,
\begin{align}
&B^z_m(\bm q)= \nonumber \\
&i\sum_{n (\neq m)} 
\frac{
\bra{\bm q,m}{\partial_{q_x}}\mathcal{H}(\bm q)\ket{\bm q,n}
\bra{\bm q,n}{\partial_{q_y}}\mathcal{H}(\bm q)\ket{\bm q,m}-{\rm c.c.}
}
{(\varepsilon_{\bm q,m}-\varepsilon_{\bm q,n})^2},
\end{align}
where $m$ and $n$ are indices of conduction-electron bands, and $\mathcal{H}(\bm q)$ is the Fourier-transformed Hamiltonian~\cite{TKNN1982}. The Fermi surface in Fig.~\ref{Fig04}(a) apparently lacks the $C_2$ point-group symmetry to the M$^{''}$-$\Gamma$-M$^{'''}$ axis reflecting the symmetry of SkX spin structure, and thus the original $D_{6h}$ symmetry of the triangular-lattice system is reduced to the $C_{3h}$ symmetry, which leads to inequivalence between the K$^{''}$-$\Gamma$ and K$^{'''}$-$\Gamma$ lines. The observed nonreciprocal nature of the spin excitation is attributable to this asymmetry. In fact, the $C_2$ symmetry to K$^{''}$-$\Gamma$-K$^{'''}$ axis is also absent, but the related asymmetry is so small and thus invisible in the Fermi surface in Fig.~\ref{Fig04}(a). On the contrary, the absence of both $C_2$ symmetries is clearly seen in the Berry curvature $B^z_m(\bm q)$ in Fig.~\ref{Fig04}(b). The predicted nonreciprocal nature of the Fermi surface might be observed by the angle-resolved photoemission spectroscopies in real skyrmionic materials (see Supplementary Material). Note that a certain spin-wave mode of skyrmion-tubes in chiral magnets exhibit nonreciprocal propagations~\cite{Seki2020,Kravchuk2020} owing to the DMI, but its origin and behaviors are distinct from those in the present centrosymmetric system.

In summary, we have theoretically studied characteristic spin and charge excitations for SkX phases in centrosymmetric itinerant Kondo-lattice magnets. We have discovered clear segregation of spin and charge excitations with three linearly dispersive Goldstone modes in the spin channel and one quadratically dispersive pseudo-Goldstone mode in the charge channel in the zero-field SkX phase. We have also revealed the nonreciprocal nature of spin excitations in another SkX phase under a magnetic field. Discoveries of itinerant skyrmion-hosting materials with centrosymmetric crystals have been successively reported recently. We expect that the present work will accelerate the research on physics of magnetic topology.

R.E. and R.P. thank Yutaka Akagi for fruitful discussions. R.E. also thanks Shinji Watanabe for fruitful discussions. This work was supported by JSPS KAKENHI (20H00337, 19H00864, and 16H06345), JST CREST (JPMJCR20T1). R.P. is supported by MEXT ``Program for Promoting Researches on the Supercomputer Fugaku" (JPMXP1020200104) and JSPS KAKENHI ``Quantum Liquid Crystals" (JP19H05825). Numerical calculations were carried out at the Supercomputer Center in Institute for Solid State Physics, University of Tokyo.

\end{document}